\documentclass[reprint,superscriptaddress, amsmath,amssymb, aps]{revtex4-2}

\usepackage{xcolor}
\usepackage{graphicx}
\usepackage{dcolumn}
\usepackage{bm}

\begin{document}

\title{Sampling rare trajectories using stochastic bridges}

\author{Javier Aguilar}
\author{Joseph W. Baron}
\author{Tobias Galla}
\author{Ra\'ul Toral}

\affiliation{Instituto de F\'{\i}sica Interdisciplinar y Sistemas Complejos IFISC (CSIC-UIB), Campus UIB, 07122 Palma de Mallorca, Spain. }

\date{\today}

\begin{abstract}
 The numerical quantification of the statistics of rare events in stochastic processes is a challenging computational problem. We present a sampling method that constructs an ensemble of stochastic trajectories that are constrained to have fixed start and end points (so-called stochastic bridges). We then show that by carefully choosing a set of such bridges and assigning an appropriate statistical weight to each bridge, one can focus more processing power on the rare events of a target stochastic process while faithfully preserving the statistics of these rate trajectories. Further, we also compare the stochastic bridges produced using our method to the Wentzel-Kramers-Brillouin (WKB) optimal paths of the target process, derived in the limit of low noise. We see that the paths produced using our method, encoding the full statistics of the process, collapse onto the WKB optimal path as the level of noise is reduced. We propose that our method can be used to judge the accuracy of the WKB approximation at finite levels of noise.
\end{abstract}

\maketitle

{\em Introduction.} The most uncommonly occurring events in stochastic systems are often the most consequential. Instances where this unlikely-yet-important combination occurs include fade-outs of epidemics \cite{assaf2017wkb,kamenev2008extinction}, the extinction of species in ecology \cite{mobilia2010fixation,kessler2007extinction}, the dynamics of biological switches \cite{li2014,alamilla2012reconstructing,Wang2011,onuchic1997theory,warren2005,bhattacharyya2020stochastic},  the escape of a Brownian particle from a double-well potential \cite{talkner1988transition,simon1992escape}, large fluctuations in chemical reactions \cite{dykman1994large} and the detection or prediction of rare natural disasters such as earthquakes, storms or heavy rains ~\cite{gabrielov2000colliding,frei2001detection}. The broad range of these applications justifies the considerable recent effort expended on developing sampling algorithms for rare events in models of stochastic phenomena \cite{malik2020rare,bouchet2019rare,khasin2011control,hurtado2020building,carollo2018making}.

Rare events can often be conceived of as paths in phase space connecting long-lived states. A number of different approaches exist to generate these transition paths for a given system. The celebrated Wentzel-Kramers-Brillouin (WKB) method, for example, is not only used to compute quasi-stationary distributions and non-equilibrium  landscapes~\cite{wentzell1998random,Wang2011,bhattacharyya2020stochastic,wang2008potential,wang2010kinetic,ye2021landscape}, but it also delivers paths describing rare events. This approach relies on a saddle-point approximation in the limit of weak noise. As a consequence, the WKB instanton provides information about the most likely path by which a systems transits from one long-lived state to another \cite{kessler2007extinction,assaf2017wkb,ashcroft2016wkb,Dembo1998,Heymann2008,grafke2015instanton,grafke2019numerical}. Little can be learned from the WKB approach about the statistics of transition paths in stochastic systems with finite noise.

Transition path sampling algorithms ~\cite{dellago1998transition,dellago2009transition,bolhuis20113} and forward flux techniques ~\cite{dellago2009transition,Allen2009,berryman2010sampling} to sample rare events account for stochasticity with finite amplitude. Transition path sampling starts from an initial trajectory connecting two long lived states and then uses a Metropolis scheme to systematically update this path. Trajectories can then be sampled to faithfully reflect the statistics of the system at hand. Forward flux techniques divide phase space (or a reduced reaction coordinate space) into patches. Transition paths are then constructed as a sequence of small segments connecting these patches. Other techniques such as so-called weighted ensemble methods \cite{huber1996weighted,Donovan2016} also rely on a segmentation of phase space. 

While these powerful tools are widely used to sample rare events, each of these approaches also has limitations. Transition path sampling methods, for example, require detailed balance \cite{bolhuis2010trajectory}, and forward flux algorithms are known to draw statistically biased trajectories \cite{van2012dynamical}. Recent work has focused on combining the strengths of these two strategies~\cite{buijsman2020transition}.

In this work we use so-called `stochastic bridges' \cite{Gasbarra2007},  which pass  through specified start and end points  by construction,  to quantify the statistics of rare trajectories. Gaussian bridges (bridges for Gaussian processes), a subclass of stochastic bridges, are used in physics \cite{benichou2016joint,delorme2016extreme}, finance \cite{mori2019time,mori2020distribution} and information processing \cite{Menguturk2018}. Further, Langevin bridges have been used to generate trajectories connecting long-lived states of stochastic differential equations (SDEs)~\cite{majumdar2015effective,orland2011generating}.

The method we present here can be summarised as follows: For a given target stochastic process, we define a bespoke stochastic bridge process. We show that the statistics of the target process can be recovered by associating a statistical weight with each stochastic bridge. This allows us to dedicate more computational effort to rare trajectories without introducing bias or interdependence. The method is flexible; the target process is fully general, the time and state space of the target process can each be discrete or continuous, detailed balance is not required, and no small-noise approximation is made. 

We show further that the stochastic bridges our method produces provide the full statistics of the ensemble of transition paths between long-lived states of the target process. This allows one to sample fluctuations around the WKB instanton, and thus to judge if the WKB approximation scheme is accurate at various levels of noise.

{\em Motivation and simple example.} We first consider a Markov process in discrete time, $t=0,1,2,\dots,T$, with a discrete set of states which we label $x_t$. The process is defined by the transition probabilities $W^t_{x\rightarrow y}=P(x_{t+1}=y|x_t=x)$ and a probability distribution $P_{0}\left(x_0\right)$ for the initial state $x_0$. As indicated by the superscript $t$ we allow for an explicit time dependence of the transition probabilities. We will refer to the ordered sequence of states visited in a realisation of the process as a {\em path}. We write this as $\mathcal{T}=(x_0, x_1, x_2,\dots, x_{_T})$, noting that the same state can be visited multiple times along a path. 
The probability to observe a particular path $\mathcal{T}$ is 
\begin{eqnarray}
\mathcal{P}\left(\mathcal{T}\right)=P_{0}\left(x_{0}\right)W^0_{x_{0}\rightarrow x_{1}}W^1_{x_{1}\rightarrow x_{2}}\cdots W^{T-1}_{x_{_{T-1}}\rightarrow x_{_T}}.
\label{eq:Prob_T}
\end{eqnarray}
These probabilities fully characterise the process. For example, the probability of finding the system in state $x$ at time $t$ is
\begin{eqnarray}
P\left(x,t\right)=\sum_{\mathcal{T}}\mathcal{P}\left(\mathcal{T}\right)\delta_{x,x_{_t}},
\label{eq:Prob_of_being_in_x_at_t}
\end{eqnarray}
where $\delta_{x,y}$ is the Kronecker delta.
A random walk on the set of non-negative integers is a simple example of such a process. We assume that the walker departs from a fixed state $x_{0}$. The transition rates are $W_{x\rightarrow x+1}=p$ and $W_{x\rightarrow x-1}=1-p$ ($0\leq p \leq 1$). The random walk is biased when $p\neq 1/2$. We assume a reflecting boundary at $x=0$, i.e., $W_{0\to 1}=1$.

\begin{figure}[t]
 \centering
 \includegraphics[width=0.4\textwidth]{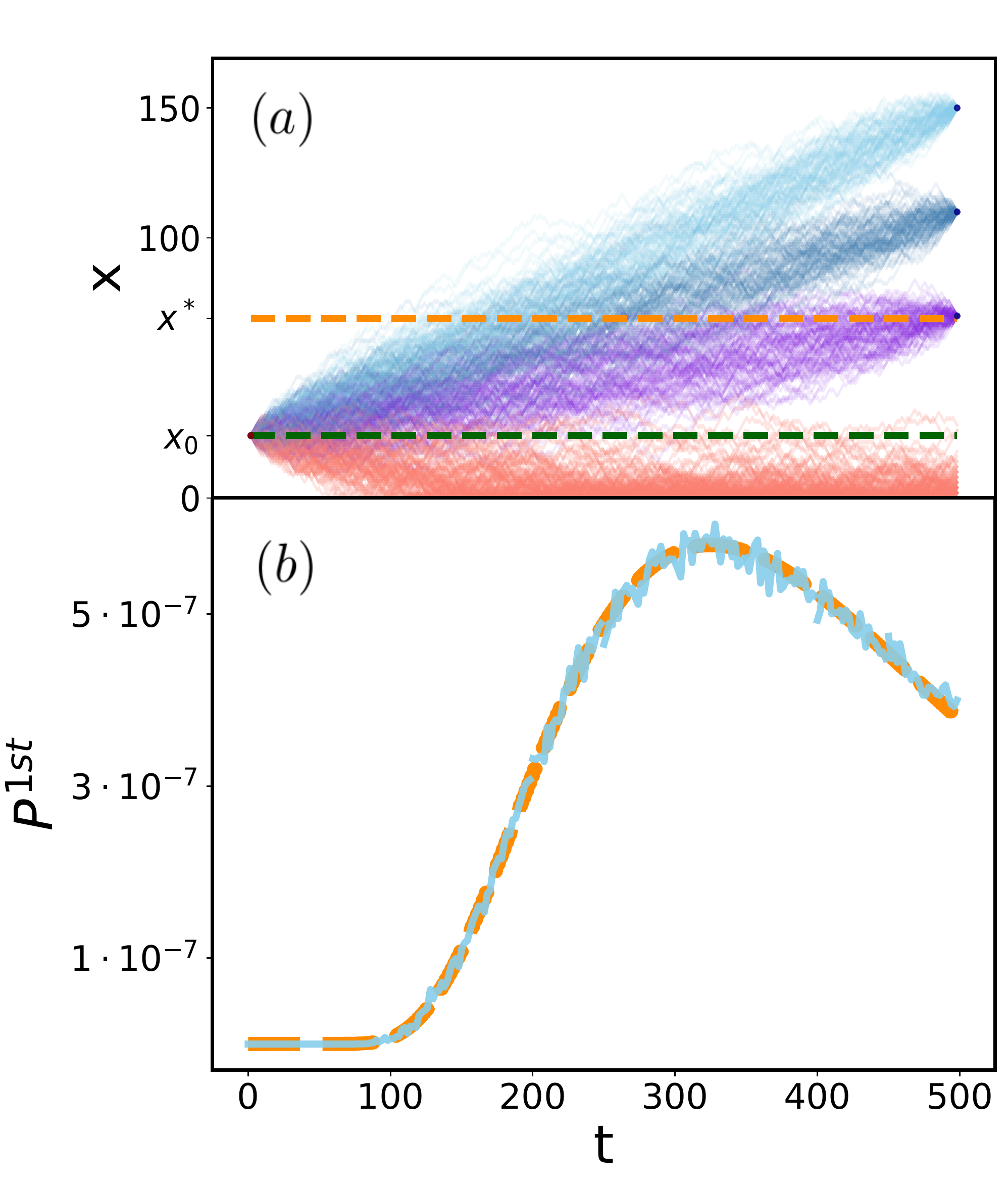}
 \caption{(a) Red (bottom) lines show trajectories generated from simulations of the biased random walk described in the text (for $p=0.45$), departing from $x_0=25$. We use reflecting boundary conditions ($W_{0 \rightarrow 1}=1$). None of these trajectories were found to cross $x^*=70$ before time $t=500$. Lines in the upper part of panel (a) show stochastic bridges with fixed initial state $x_{0}=25$ and final states $x_{_T}=150$, $x_{_T}=130$ and $x_{_T}=70$ respectively (top to bottom). (b) Distribution of first-passage times at $x=x^*$ computed from our approach (See Section \ref{Ap-BRW} of SM for details). Also shown (dashed line) is the analytical result from \cite{kager2011hitting}.}
 \label{fig:BRW}
\end{figure}

Typical paths of this process  are illustrated in  Fig.~\ref{fig:BRW}(a). In the figure the random walk is biased towards lower integers. The probability that a realisation terminates at a state $x_{_T}>x_{0}$ is then low, in particular when $x_{_T}$ is much larger than $x_{0}$ and/or when $p$ is much smaller than $1/2$. It is then difficult to sample paths ending at values $x_{_T}>x_0$ in direct simulations of the biased walk.

{\em{Method.}} Our strategy for sampling paths of a given target dynamics connecting specific start and end points, $x_0$ and $x_{_T}$ respectively, is based on an `associated bridge process' operating backwards in time. That is, paths of this bridge process are generated from $x_{T}$ to $x_0$. We first describe this for the case of discrete time and discrete states, generalisations are discussed below. 

We define the associated bridge process via transition probabilities $\tilde{W}^t_{x\leftarrow y}=P(x_t=x|x_{t+1}=y)$. That is to say $\tilde{W}^t_{x\leftarrow y}$ is the probability that the target process was at $x_t$ at time $t$, given that state $y$ is visited at time $t+1$. To define the associated bridge process we also need to specify a probability distribution $\tilde{P}_{T}\left(x\right)$ for the state $x_{_T}$. We require of this distribution that it assigns no probability to states that are not accessible as final states of the target process, i.e., $\tilde{P}_{T}\left(x\right)$ must be zero if $P\left(x,T\right)= 0$. In the random walk in Fig.~\ref{fig:BRW} for example, states with $|x_{_T}-x_0|>T$ cannot be reached.  

 By virtue of Bayes' theorem we have
\begin{eqnarray}\label{eq:Backward_transition_prob}
\tilde{W}^t_{x\leftarrow y}=W^t_{x\rightarrow y}\frac{P\left(x,t\right)}{P\left(y,t+1\right)}.
\end{eqnarray}
As a consequence the probabilities $\tilde{W}^t_{x\leftarrow y}$  will in general be time-dependent, even if the transition probabilities of the original model do not depend on time.

The rates $\tilde{W}^t_{x\leftarrow y}$ defined in Eq.~(\ref{eq:Backward_transition_prob}) might seem reminiscent of the Doob transform of the target process \cite{doob, levin2017markov,majumdar2015effective,Chetrite2015}, see   Sec.~\ref{Ap-Doob} of the Supplemental Material (SM) for further details. However, it is important to note that there are significant differences between the two. In particular, the fact that we define the associated bridge process backwards in time means that the probabilities $P(x,t)$ and $P(y,t+1)$ on the right hand side Eq.~(\ref{eq:Backward_transition_prob}) are not conditioned on any final state $x_{_T}$ , so  improving the overall efficacy of the sampling method (see Sec.~\ref{Ap-Doob} of the SM).

We now focus on a fixed path $\mathcal T=(x_0,\dots, x_{_T})$ of the target process. The path can also occur in the associated bridge process where it is traversed starting at $x_{_T}$ and ending at $x_0$. The probability to observe path ${\cal T}$ in the associated bridge process is
\begin{eqnarray}
\tilde{\mathcal{P}}(\mathcal{T})=\tilde P_T(x_{_T})\tilde{W}^T_{x_{_{T-1}}\leftarrow x_{_T}} \cdots \tilde W^{1}_{x_0\leftarrow x_1} . 
\label{eq:Prob_T_B}
\end{eqnarray}
Combining Eqs.~\eqref{eq:Prob_T}, (\ref{eq:Backward_transition_prob}) and (\ref{eq:Prob_T_B}) we find a relation between the probabilities of finding path ${\cal T}$ in the original and associated processes respectively,
\begin{eqnarray}
\mathcal{P}(\mathcal{T})=\tilde{\cal P}(\mathcal{T})\frac{P\left(x_{_T},T\right)}{\tilde{P}_T\left(x_{_T}\right)}.
\label{eq:key_eq_general}
\end{eqnarray}
Eqs.~(\ref{eq:Backward_transition_prob}) and (\ref{eq:key_eq_general}) are the key components of our approach. We use the process defined by Eq.~(\ref{eq:Backward_transition_prob}) to generate paths ${\mathcal T}$, i.e., we sample from $\tilde P({\cal T})$. Using Eq.~(\ref{eq:key_eq_general}) we then read off the probability with which each sample path occurs in the target process.

If we choose $\tilde{P}_T(x)=P(x,T)$ then ${\cal P}({\cal T})=\tilde {\cal P}({\cal T})$ and paths that are rare in the target process will also be rare in the associated bridge process. Eqs. (\ref{eq:Backward_transition_prob}) and (\ref{eq:key_eq_general}) then do not constitute an efficient sampling method for rare paths. If, on the other hand, we choose $P_{0}(x)=\delta_{x,x_0}$ and $\tilde{P}_{T}(x)=\delta_{x,x_{_T}}$, the trajectories generated from the process in Eq.~(\ref{eq:Backward_transition_prob}) will connect $x_{_T}$ and $x_0$, i.e., they are realisations of a stochastic bridge.  For such paths Eq.~\eqref{eq:key_eq_general}  then reduces to
\begin{eqnarray}
\mathcal{P}(\mathcal{T})=\tilde{\mathcal{P}}(\mathcal{T})P\left(x_{_T},T\right).
\label{eq:key_eq}
\end{eqnarray}
While the most demanding part of the procedure in terms of computing time is the calculation of $P(x,t)$ for $t=0,1,\dots,T$, this can usually be obtained efficiently. In simple examples $P(x,t)$ can be found by integrating the master equation (see e.g. \cite{dickman2002numerical}). Established numerical methods are also available for systems with continuous states  ~\cite{risken1996fokker,Kromer2013}. The complexity can be reduced further for escape paths from long-lived states. The distribution $P$ in Eq.~(\ref{eq:Backward_transition_prob}) can then be replaced by the time-independent quasi-stationary distribution $P^\text{QS}$ describing the metastable state. The transition rates for the associated process then become time-independent, $
\tilde{W}_{x\leftarrow y}=W_{x\rightarrow y}P^\text{QS}\left(x\right)/P^\text{QS}\left(y\right)$. For some models the quasi-stationary distribution can be approximated analytically, for example with the WKB method (see e.g. ~\cite{dickman2002numerical,ashcroft2016wkb,meerson2009wkb,mobilia2010fixation,assaf2010extinction}, and also Section \ref{Ap-SIS} of SM ).

We now return to the example of a biased random walk. Fig.~\ref{fig:BRW}(a) shows three ensembles of bridges, all starting at a fixed value of $x_0$ at $t=0$ and each ensemble ending at a different choice of $x_{_T}$ at time $t=T$. These paths were generated from the associated process in Eq.~(\ref{eq:Backward_transition_prob}), where we have chosen  $\tilde P(x_{_T})$ as delta functions at the desired endpoints. We highlight that not all trajectories in these different ensembles are equally likely in the target random walk process. Instead their probability weights are given by Eq.~(\ref{eq:key_eq_general}).

For a fixed final time $T$ and choice of $x_{_T}$ we can determine whether the generated trajectories have crossed a fixed $x^*$ ($x_{0}<x^*<x_{_T}$) by time $T$. For each trajectory that has crossed $x^*$ we can record the first crossing time. Repeating the process for different values of $x_{_T}$ and weighing trajectories according to Eq.~(\ref{eq:key_eq_general}) we then obtain the distribution of first-passage times through $x^*$ (see Sec.~\ref{Ap-BRW} of the SM for details). For the example of the random walk, this distribution can also be calculated analytically \cite{kager2011hitting}. A comparison of our simulations with these predictions is shown in Fig.~\ref{fig:BRW}(b), confirming the validity of our sampling method.

{\em Continuous-time processes.} The method can also be used when time and/or the state space of the target process are continuous. The only modification is an adjustment of the expression in Eq.~(\ref{eq:Backward_transition_prob}). If time is continuous one has  \begin{eqnarray}
\tilde{\omega}^t_{x\leftarrow y}&=&\omega^t_{x\rightarrow y}\frac{P\left(x,t\right)}{P\left(y,t+dt\right)}\nonumber \\
&=&\omega^t_{x\rightarrow y}\frac{P\left(x,t\right)}{P\left(y,t\right)}+{\cal O}(dt),
\label{eq:Backward_transition_rates_continuous_time}
\end{eqnarray}
where we have written $\omega_{x\to y}^t$ for the transition rates of the target dynamics, and $\tilde{\omega}^t_{x\leftarrow y}$ for those of the associated process.
For discrete states a process described by time-dependent rates $\tilde{\omega}^t_{x\leftarrow y}$ can be simulated for example using Lewis' thinning algorithm~\cite{lewis1979simulation,Ogata}. If these rates have no explicit time dependence, then the standard Gillespie method is sufficient. If the focus is on escape phenomena from metastable states, then the distribution $P(\cdot, t)$ on the right-hand side Eq.~(\ref{eq:Backward_transition_rates_continuous_time}) can again be replaced with the corresponding quasi-stationary distribution of the target process. We also note that $\tilde{\omega}^t_{x\leftarrow y}$ is a Gaussian bridge process \cite{Gasbarra2007,Menguturk2018,delorme2016extreme} when the original process is Gaussian (for details see Sec.~\ref{Ap-Langevin} of the SM). 

{\em Applications: Models of an epidemic and of a simple genetic switch.} In the context of two examples, we now compare the WKB trajectories of the target process to the associated bridge trajectories starting and ending in the same positions. The WKB optimal path of the target process is the most likely path that the system will take (in the limit of small noise) given the end points $x_0$ and $x_{_T}$.  At finite noise levels our method captures stochastic fluctuations about the WKB instanton. As the level of noise is reduced, the realisations of the associated bridge process approach the WKB trajectory.

We first focus on the extinction of an epidemic (or `fade out') in the individual-based SIS model. Conceptually this is similar to extinctions of species in ecology~\cite{bartlett1960stochastic}. The model describes $N$ individuals, of which $n$ are infected. The population evolves in continuous time via infection and recovery processes, with rates
\begin{eqnarray}
\label{eq:SIS_rates}
\omega_{n\rightarrow n+1}=\beta \frac{n(N-n)}{N}, \quad 
\omega_{n\rightarrow n-1}=\gamma n. 
\end{eqnarray}
The parameters $\beta$ and $\gamma$ characterise speed of infection and recovery respectively.

When $\beta>\gamma$, the model is known to evolve to a quasi-stationary `endemic' state in which the number of infected individuals fluctuates about $n= N(1-\gamma/\beta)$ \cite{marro2005nonequilibrium}. 
The population will remain in this metastable state until a large fluctuation drives the epidemic to extinction. The mean time to reach this absorbing state grows exponentially with the population size $N$ \cite{naasell1996quasi}. For large populations it is therefore difficult to observe extinction in direct simulations of the SIS-dynamics. Such paths can however be generated straightforwardly using the associated bridge process with rates given by Eq.~\eqref{eq:Backward_transition_rates_continuous_time}. To evaluate these rates we use the analytical solution for the quasi-stationary distribution in $P^\text{QS}(n)$~\cite{naasell1996quasi} (see also Section \ref{Ap-SIS} of the SM). 

\begin{figure}[h]
 \centering
 \includegraphics[width=0.35\textwidth]{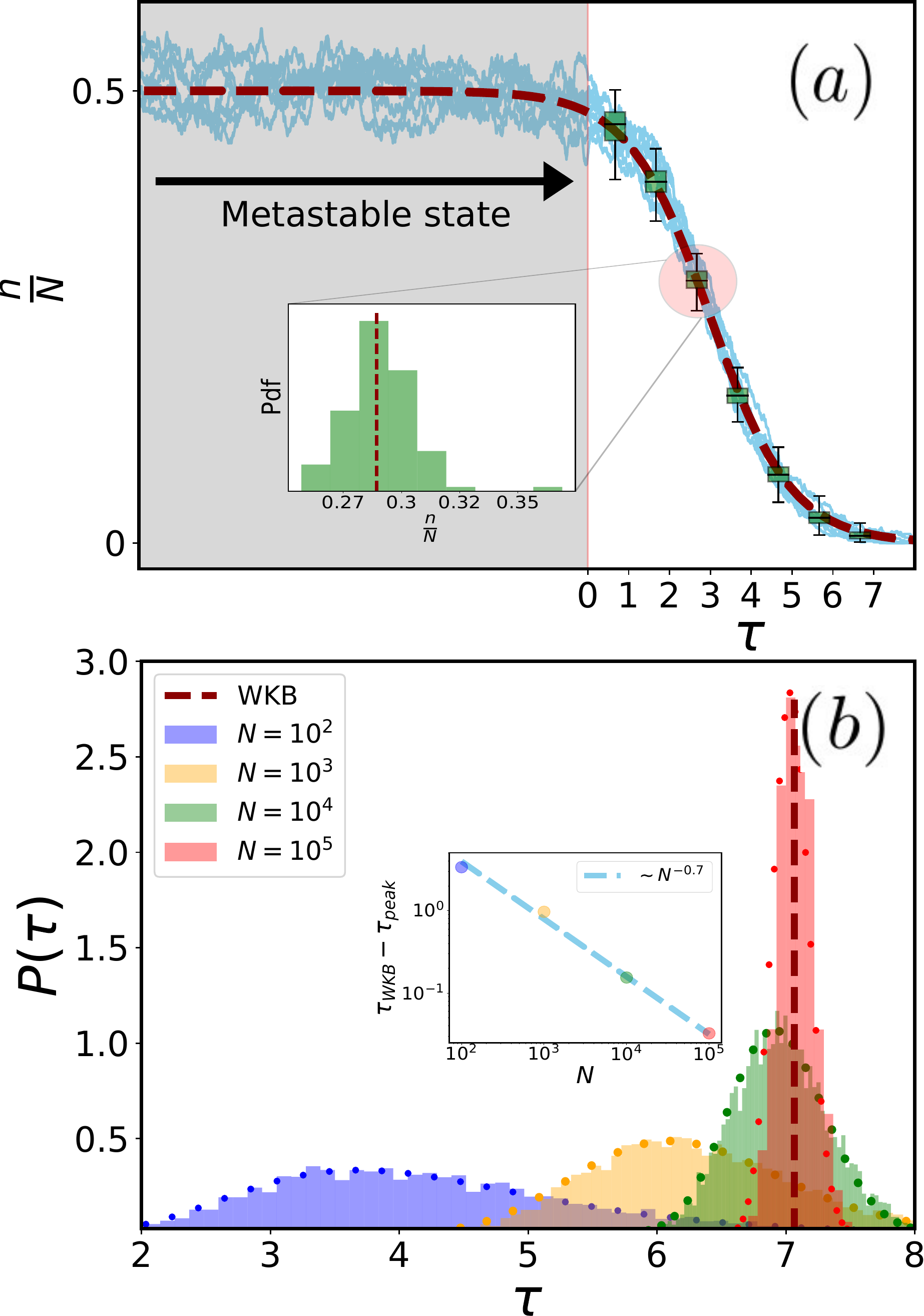}
 \caption{Extinction paths for the SIS model ($\beta=2, \gamma=1$). (a) Paths leading to extinction from a common starting point $\frac{n}{N}=1-\frac{\gamma}{\beta}=0.5$ to $n=0$ for $N=10^3$. The WKB instanton is shown as a dashed line. The time $\tau=0$ corresponds to the point where the WKB instanton crosses $n/N = 0.48$. Boxes indicate the median and first quartiles, and error bars the observed range of the ensemble of stochastic paths. The inset shows the distribution of $n/N$ at time $\tau=2.8$. (b) Distribution of transition times for extinction trajectories from the quasi-stationary state towards the absorbing state (see Sec.~\ref{Ap-SIS} of the SM for details). Dots show fits to log-normal distributions. The inset shows that the modes $\tau_{\text{peak}}$ of these fits approach the value predicted from the WKB instanton, with $|\tau_{_\text{WKB}}-\tau_{\text{peak}}|\sim N^{-0.7}$.  }
 \label{fig:SIS}
\end{figure}

An ensemble of extinction paths is shown in Fig.~\ref{fig:SIS}(a), along with the WKB instanton to extinction \cite{ashcroft2016wkb,assaf2010extinction}. As is illustrated in Fig.~\ref{fig:SIS}(a), the extinction paths at finite $N$ fluctuate about the WKB instanton. We also show the distribution of transition times towards the infection-free state in Fig.~\ref{fig:SIS}(b). This time scale characterises the duration of the transition towards extinction once the system has left the endemic state, and is not to be confused with the lifetime of the metastable state itself \cite{wentzell1998random,ashcroft2016wkb,assaf2010extinction,mobilia2010fixation,meerson2009wkb} (see also Sec.~\ref{Ap-SIS} of the SM). This distribution is relatively broad for small populations, but becomes more and more concentrated on the WKB estimate for larger $N$. 

\begin{figure}[h]
 \centering
 \includegraphics[width=0.48\textwidth]{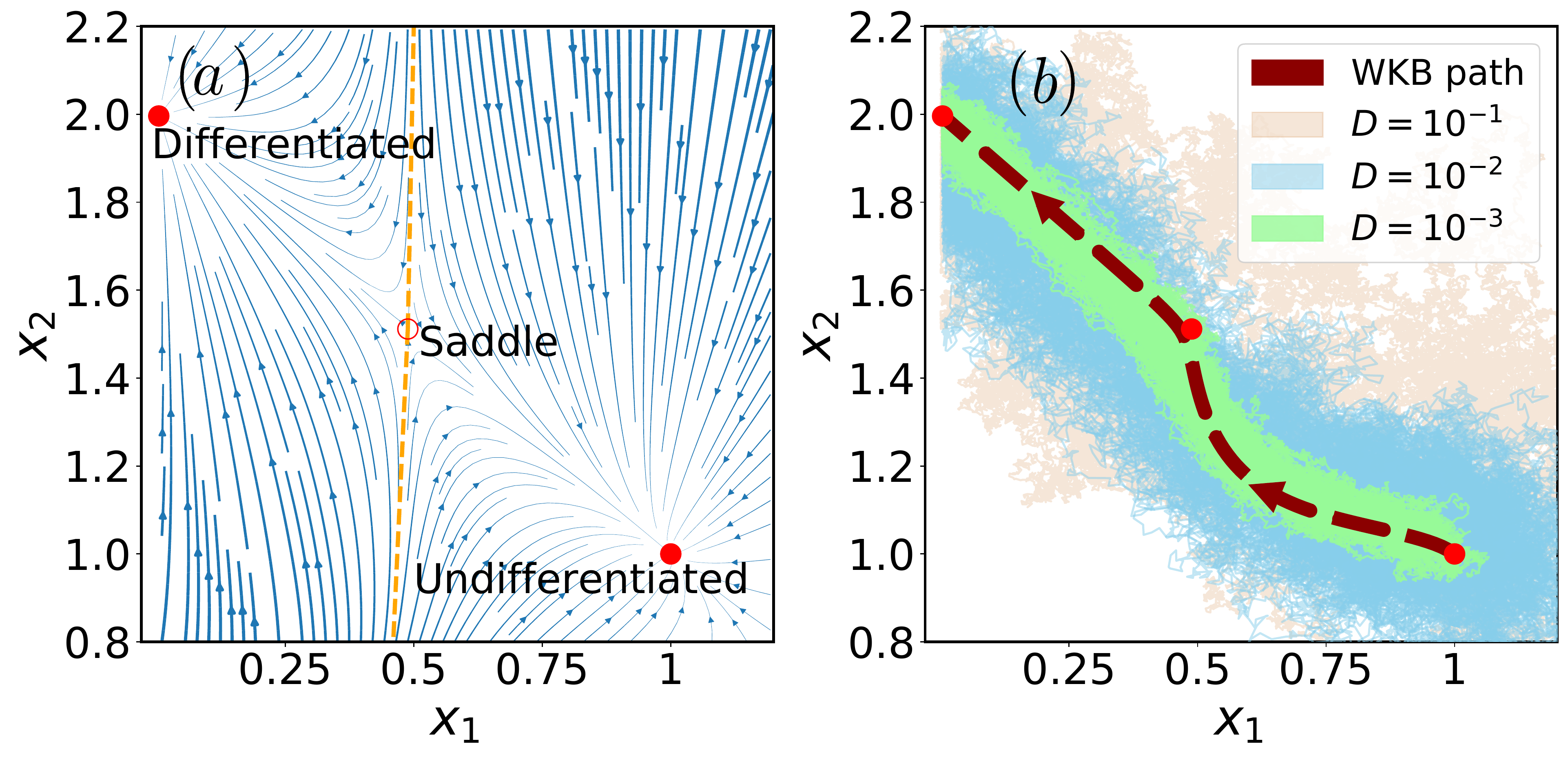}
 \caption{(a) Phase portrait of the system in Eqs.~(\ref{eq:Waddington_additive_noise}) for $D=0$. Arrows indicate the deterministic flow. Due to the $x_1\leftrightarrow x_2$ symmetry we only show one of the two fixed points describing differentiated states, and one saddle point. The corresponding separatrix is indicated as a dashed line. (b) Stochastic transition paths from the undifferentiated to one the differentiated states obtained with our sampling method. The dashed line is the WKB instanton.}
 \label{fig:cell}
\end{figure}

As a second example, we focus on a model of cell differentiation discussed in \cite{Wang2011a}. The two real variables $x_1\geq 0$ and $x_2\geq 0$ in the model describe protein concentrations, governed by the SDEs
\begin{eqnarray}
 \dot{x}_{1}&=&\frac{x_{1}^{n}}{S^{n}+x_{1}^{n}}+\frac{S^{n}}{S^{n}+x_{2}^{n}}- \ x_1 +\sqrt{2D}\xi_{1}(t),\nonumber \\
 \dot{x}_{2}&=&\frac{x_{2}^{n}}{S^{n}+x_{2}^{n}}+\frac{S^{n}}{S^{n}+x_{1}^{n}}- \ x_2 +\sqrt{2D}\xi_{2}(t),\label{eq:Waddington_additive_noise}
\end{eqnarray}
where $\xi_1(t)$ and $\xi_2(t)$ are Gaussian noise variables with mean zero and $\langle \xi_i(t)\xi_j(t')\rangle=\delta_{ij}\delta(t-t')$. The noise describes effects external to the gene circuit and its strength is governed by the model parameter $D\geq 0$. The deterministic terms on the right-hand side of Eqs.~(\ref{eq:Waddington_additive_noise}) represent self-activation, mutual inhibition and degradation respectively \cite{wang2010kinetic}. Through the rest of the work, the model parameters $n=4$ and $S=0.5$ are fixed.

In the deterministic limit ($D=0$) Eqs.~(\ref{eq:Waddington_additive_noise}) have three stable fixed points, (i) one with $x_1=x_2=1$, (ii) one with $x_1>x_2$, and (iii) a third one obtained from the second by exchanging $x_1$ and $x_2$.  The first fixed point describes an undifferentiated cell state, the other other two differentiated states  \cite{Wang2011}. The deterministic model also has two saddle points on the separatrices between the basins of attraction of the fixed points. The resulting phase portrait is shown in Fig.~\ref{fig:cell}(a).

For $D>0$, noise-driven transitions from the undifferentiated to either one of the differentiated states become possible. Similar to the SIS model the typical escape time grows exponentially with the inverse noise strength~\cite{hanggi1990reaction}. For small noise amplitudes, numerical integration of Eqs.~(\ref{eq:Waddington_additive_noise}) is therefore unlikely to generate transition paths within realistic computing times.  

To sample these rare paths  we first discretize time using an Euler-Maruyama scheme~\cite{kloeden1992stochastic, toral2014stochastic}. The resulting process is described by Gaussian transition rates. These are then used in Eq.~(\ref{eq:Backward_transition_rates_continuous_time}) together with the quasi-stationary distribution describing the undifferentiated state. This quasi-stationary distribution in turn is obtained from a numerical integration of the Fokker-Planck equation for the dynamics in Eqs.~(\ref{eq:Waddington_additive_noise}), see Sec.~\ref{AP-FP} of the SM. As a result of this procedure we obtain an ensemble of trajectories starting from the undifferentiated state and ending in the differentiated state $x_2>x_1$ (see also Sec.~\ref{Ap-random_number_generator} in the SM).
In Fig.~\ref{fig:cell}(b) we show ensembles of transition paths for different noise amplitudes. As demonstrated in the figure the stochastic trajectories approach the WKB instanton as $D\to 0$. 

The level of noise at which the WKB approximation is useful is very much dependent on the application in question. The comparison between the stochastic bridges generated with our method and the WKB instantons allows us to appraise the accuracy of WKB results, such as stationary probability distributions or transition rates between basins of attraction. 

{\em Conclusions:} In this Letter, we have presented a method to sample rare trajectories. The core component of the approach is a process in reverse time that generates stochastic bridges connecting desired start and end points. Using Eq. (\ref{eq:key_eq_general}) we can then calculate the probability with which these paths occur in the target process under consideration. This allows us to quantify the statistics of rare events such as first passage times. Our approach does not require the noise in the model to be weak, and it generates uncorrelated and unbiased transition paths. 

Traditional WKB methods provide information about the most likely path connecting two states, and allow one to calculate typical quantities characterising these transitions, e.g. mean first passage or transition times. Our approach goes beyond this, and delivers an ensemble of transition paths along with their statistical weights. This enables us to obtain entire distributions of first-passage times or other characteristics in simulations.

We envisage that the method that we have developed will have applications in myriad systems where sampling rare events is important. We imagine that it can also be used as a numerical aid to intuit when the WKB method will be accurate and useful. The approach presented here can also be extended to sample stochastic trajectories constrained to pass through more than two desired points.

\begin{acknowledgments}
We thank Pere Colet, Tobias Grafke, Jin Wang, Horacio Wio, and Kun Zhang for useful discussions. Partial financial support has been received from the Agencia Estatal de Investigaci\'on (AEI, MCI, Spain) and Fondo Europeo de Desarrollo Regional (FEDER, UE) under Project PACSS (RTI2018-093732-B-C21/C22), and the Mar{\'\i}a de Maeztu Program for units of Excellence in R\&D, grant MDM-2017-0711 funded by MCIN/AEI/10.13039/501100011033.
\end{acknowledgments}

\newpage

\onecolumngrid
 \clearpage
 
 \begin{center}
 \Large{--- Supplemental Material ---}
\end{center}
\setcounter{figure}{0}
\setcounter{equation}{0}
\renewcommand{\thesection}{S\arabic{section}} 		
\fontsize{12}{14.5}\selectfont
\section{Relation with Doob's $h$-transform}\label{Ap-Doob}
Despite a superficial resemblence, Eq.~(\ref{eq:Backward_transition_prob}) is different from Doob's so-called  h-transform \cite{levin2017markov,sarkka2019applied}. In this section we provide details regarding these differences. We also explain in what sense the method based on the associated process in reverse time [Eq.~\eqref{eq:Backward_transition_prob} in the main paper] is more efficient for our purposes than the use of the Doob's h-transform.

\subsection{Doob's $h$-transform}
We first focus on discrete-time processes with transition probabilities $W^t_{x \rightarrow y}$. For a given function function $h(\cdot,\cdot)$ Doob's h-transform \cite{doob,sarkka2019applied,Chetrite2015,Chung2005} is then a process defined by the transition rates
\begin{equation}
    \label{eq:AP_Doob_general_Doob_transform}
    \hat{W}^t_{x\rightarrow y}=W^t_{x\rightarrow y}\frac{h(y,t+1)}{h(x,t)}.
\end{equation}
The positive function $h$ can be chosen arbitrarily provided that it fulfills
\begin{equation}
    \label{eq:AP_Doob_normalization_relation_time}
    h(x,t)=\sum_y \ W^t_{x\rightarrow y} \ h(y,t+1).
\end{equation}
This condition ensures the normalisation $\sum_{y}\hat{W}^t_{x\rightarrow y}=1$.

If time is continuous a very similar definition applies
\begin{equation}
    \label{eq:AP_Doob_general_Doob_transform_time}
    \hat{w}^t_{x\rightarrow y}=w^t_{x\rightarrow y}\frac{h(y,t)}{h(x,t)},
\end{equation}
with the condition
\begin{equation}
    \label{eq:AP_Doob_normalization_relation}
    h(x,t)=\int dy \ w^t_{x\rightarrow y}  h(y,t).
\end{equation}

\subsection{Construction of stochastic bridges from the Doob transform}

A suitable choice of the function $h(x,t)$ in Eq.~(\ref{eq:AP_Doob_general_Doob_transform}) allows one to generate conditioned Markov processes. Consider for example 
\begin{equation}\label{eq:def_h}
h(x,t)=P(x_{_T}|x_t=x),
\end{equation}
where $P(x_{_T}|x_t=x)$ is the probability that a trajectory of the original process visits $x_{_T}$ at time $T$, given that it was at $x$ at time $t$. We then have Eq.~(\ref{eq:AP_Doob_general_Doob_transform}) as follows,
\begin{eqnarray}\label{eq:AP_Doob_stochastic_bridge}
\hat{W}^t_{x\rightarrow y}= W^t_{x\rightarrow y}\frac{P(x_{_T}|x_{t+1}=y)}{P(x_{_T}|x_t=x)}
=P(x_{t+1}=y|x_t=x,x_{_T}).
\end{eqnarray}
We have used the definition of conditional probabilities, and the fact that  $P(x_{_T}|x_t=x,x_{t+1}=y)=P(x_{_T}|x_{t+1}=y)$ for Markov processes.

The last expression in Eq.~(\ref{eq:AP_Doob_stochastic_bridge}) indicates that $\hat{W}^t_{x\rightarrow y}$ is the probability that a trajectory of the original process ultimately arriving at $x_{_T}$ hops from to $y$ in the next step if it is at $x$ at time $t$. Therefore, the $\hat{W}^t_{x\rightarrow y}$ are the transition rates one would obtain from the ensemble of trajectories of the original process that end at $x_{_T}$.

The transition rate $\hat{W}^t_{x\rightarrow y}$ is non-zero only when $P(x_{_T}|x_{t+1}=y)>0$. This means that at any time $t$ the process defined by these rates can only jump to a state $y$ if the desired final state $x_{_T}$ can be reached from $y$ in the remaining time. This statement hold in particular at time $t=T-1$. Therefore all trajectories of the process must end in $x_{_T}$.

\subsection{Relation to sampling method in the main paper}

The transition rates $\hat W_{x\leftarrow y}$ defined in Eq.~(\ref{eq:Backward_transition_prob}) of the main paper are different to those of the Doob transform $\hat{W}^t_{x\rightarrow y}$ defined in Eq.~(\ref{eq:AP_Doob_stochastic_bridge}), but they produce an ensemble of paths with exactly the same statistics. That is, the rates $\hat{W}^t_{x\rightarrow y}$ define a process that runs forwards in time from $x_0$ to $x_{_T}$ and the rates $\hat W_{x\leftarrow y}$ define a process that runs backwards in time from $x_{_T}$ to $x_0$, but each method produces the same set of paths with equal weights. 

However, one notices that in order to compute the rates in Eq.~(\ref{eq:AP_Doob_stochastic_bridge}) and simulate one bridge (with fixed start and end points), one requires the conditional probabilities $P(x_{{}_T} \vert x_t = y )$ for all values of $y$ and a given value of $x_{{}_T}$. This set of probabilities can be obtained efficiently from the backward master equation. Eq.~(\ref{eq:Backward_transition_prob}) instead involves the conditional probabilities $P(x_{t} \vert x_0 )$ for all values of $x_t$ and a fixed value of $x_{0}$. These conditional probabilities can be obtained most efficiently from the forward master equation.

The crucial difference between the two implementations is in computational efficiency in the case where the end point $x_{{}_T}$ is not fixed, i.e. when we want to produce stochastic bridges with different end points as in Fig. \ref{fig:BRW} (a). In our algorithm, we only have to integrate the forward master equation once to obtain the probabilities $P(x_{t} \vert x_0 )$. However, if we were to use the Doob transform approach, we would have to integrate the backward master equation many times to obtain different sets of probabilities $P(x_{{}_T} \vert x_t = y )$ for each final point $x_{{}_T}$. When one wishes to sum over many final points, as we did to produce Fig. \ref{fig:BRW} b, a many-fold increase in efficiency is obtained by using our associated bridge process over the Doob transform.

\section{Recovering the statistics of the target process from the associated bridge process}\label{Ap-BRW}

\subsection{General procedure}
In Eq.~(\ref{eq:key_eq_general}) in the main text, we demonstrate how the probability of observing a particular path of the associated bridge process can be related to observing that same path in the target process. In this section, we describe how this result can be used to deduce statistics of the target process from the statistics of the associated bridge process.

We write ${\cal O}$ for an observable related to an individual path. For example, this could be the first time the path crosses a given barrier, or the largest excursion from the starting point, or indeed quantities relating to multiple points in time (e.g. the `range' of a path, i.e., the difference between maximum and minimum position attained by a random walker). We write ${\cal O}({\cal T})$ for the value of ${\cal O}$ associated with path ${\cal T}$. 

Suppose now that for each possible end state $x_{_T}$ we sample $M$ paths ${\cal T}^{(i)}_{x_{_T}}$ ($i=1,\dots,M$) ending at $x_{_T}$. 
The statistical weight of each of the paths ending at a particular value $x_{_T}$ is $P(x_T,T)/M$ by virtue of Eq.~(\ref{eq:key_eq_general}), and accounting for the fact that we have generated $M$ paths for each value of $x_{_T}$. The overall mean of ${\cal O}$ over all paths of the original system (without conditioning on a particular end state) is then obtained by summing over all possible end states $x_T$,
\begin{eqnarray}\label{eq:overallaverage}
\left\langle \mathcal{O} \right\rangle = \sum_{x_{_T}}\left[ P(x_{_T},T)\frac{1}{M}\sum_{i=1}^M {\cal O}({\cal T}^{(i)}_{x_{_T}})\right].
\end{eqnarray} 

For simplicity, we have chosen to sample the same number of trajectories $M$ from all the possible end states $x_{_T}$. It is possible to sample different numbers of paths $M_{x_{_T}}$ for different end states $x_{_T}$. Eq. (\ref{eq:overallaverage}) then becomes $\left\langle \mathcal{O} \right\rangle = \sum_{x_{_T}}\left[P(x_{_T},T)\frac{1}{M_{x_{_T}}}\sum_{i=1}^{M_{x_{_T}}} {\cal O}({\cal T}^{(i)}_{x_{_T}})\right]$.

\subsection{Application to random walk [Fig.~\ref{fig:BRW}(b) in the main manuscript]}
We now give further details of this procedure for the example in  Fig.~\ref{fig:BRW}(b), namely the first passage time distribution $P^\text{1st}(x^{*},t)$ of the biased random walk. The quantity $P^\text{1st}(x^{*},t)$ is the probability that a walker first reaches $x^*$ at time $t$, i.e., $P^\text{1st}(x^{*},t)=P \left(x(t)={x^{*}}|x(t')\neq {x^{*}} \ \forall t'<t \right)$. 

The procedure that we follow to compute $P^\text{1st}(x^{*},t)$ for a fixed $t$ is as follows.
\begin{enumerate}
    \item[1.] For a fixed value of $x_{_T}$, use the rates in Eq.~(\ref{eq:Backward_transition_prob}) to construct $M$ trajectories $\mathcal{T}^i_{x_{_T}}$ ($i=1,\dots,M$) of the random walk all ending at $x_T$. 
    \item[2.] Determine how many of these $M$ trajectories first reach state $x$ at time $t$. Call this number $N_{x_{_T}}(x^\star,t)$. 
    
    \item[3.] Repeat items 1 and 2 for all possible end states $x_{_T}$.
    \item [4.] The quantity $P^\text{1st}(x^{*},t)$ is then given by
    \begin{equation}
        P^\text{1st}(x^*,t)=\sum_{x_{_T}}P(x_{_T},T)\frac{N_{x_{_T}}(x^*,t)}{M}.
        \end{equation}
\end{enumerate}
If the random walk is run up to time $T$, then all $x_0+T+1$ final states from $x=0$ to $x=x_0+T$ are possible (recall that the state zero is reflecting for the example in Fig.~\ref{fig:BRW}). In order to limit this number at large times $T$, we introduce a reflecting boundary $W_{L\rightarrow L-1}=1$ in our simulations. Provided that $L\gg x^*$ this will not materially affect the empirical first passage time distribution through $x^*$ (in our simulations we use $x^*=70,\ L=200$). Further, we have repeated the process for multiple final times $T=250,300,350,400,450,500$, and have then combined results.

\section{Reversed Langevin equation for Gaussian processes}\label{Ap-Langevin}
In this Section we show that the method described in the main text reproduces known results for Gaussian bridges \cite{Gasbarra2007}. Consider the following linear stochastic differential equation
\begin{equation}
    \label{eq:Ap_Langevin_linear_Langevin_eq}
    \Dot{x}=a(t)x+\sqrt{2D(t)}\xi(t).
\end{equation}
where we allow for a possible time-dependence of the drift coefficient $a(t)$ and of the diffusion coefficient $D(t)$. We choose the initial condition $x(t=0)=x_0$. The distribution of $x$ at time $t$ can be obtained directly from the Fokker-Planck equation describing this process \cite{risken1996fokker}, and reads
\begin{equation}
    \label{eq:Ap_Langevin_Gaussian}
    P(x,t)=\frac{e^{-\frac{(x-\mu(t))^2}{2\sigma(t)^2}}}{\sigma(t)\sqrt{2\pi}}\equiv G\left(x;\mu(t),\sigma (t)\right),
\end{equation}
where $\mu(t) = \langle x(t)\rangle$ and $\sigma(t)^2 = \langle x(t)^2\rangle - \langle x(t)\rangle^2$ and we introduce the shorthand $G(\cdot; \mu,\sigma)$ for a Gaussian distribution with mean $\mu$ and standard deviation $\sigma$. We have $\mu(t=0)=x_0,\,\sigma(t=0)=0$.

We next discretize time into intervals of length $\Delta t$ using the Euler-Maruyama prescription \cite{toral2014stochastic},
\begin{eqnarray}
x_{t+\Delta t}=x_t+a(t)x_t\Delta t +\sqrt{2D(t)\Delta t}\Gamma_t,
\end{eqnarray}
where $\Gamma_t$ are uncorrelated Gaussian variables of mean zero and variance one. Alternatively, this scheme can be understood as the generation of a random value $x_{t+\Delta t}$ using the Gaussian transition probabilities
\begin{equation}
    \label{eq:AP_Langevin_transition_prob}
    W^t_{x_t\rightarrow x_{t+\Delta t}}=G\left(x_{t+\Delta t}; x_t +a(t)x_t\Delta t ,\sqrt{2\Delta t D(t)}\right).
\end{equation}
Inserting Eqs.~\eqref{eq:AP_Langevin_transition_prob} and \eqref{eq:Ap_Langevin_Gaussian} in Eq.~\eqref{eq:Backward_transition_prob} in the main paper we find
\begin{eqnarray}
\label{eq:AP_Langevin_reversed_transition_prob_V1}
\tilde W^t_{x_t\leftarrow x_{t+\Delta t}}&=&G\left(x_{t+\Delta t};x_{t}+a(t)x_t\Delta t ,\sqrt{2\Delta t D(t)}\right)\times \dfrac{G\left(x_t;\mu (t),\sigma (t)\right)}{G\left(x_{t+\Delta t};\mu  (t+\Delta t),\sigma (t+\Delta t)\right)}.
\end{eqnarray}
The right-hand side is the exponential of a quadratic polynomial in the variable $x_t$ and hence $\tilde W^t_{x_t\leftarrow x_{t+\Delta t}}$, has a Gaussian functional form. We note that $\mu(t)$ and $\mu(t+\Delta t)$ feature on the right-hand side, and similarly $\sigma(t)$ and $\sigma(t+\Delta t)$. Expanding the coefficients of the polynomial in powers of $\Delta t$ one finds for the reversed transition probabilities:

\begin{eqnarray}
\label{eq:AP_Langevin_reversed_transition_prob_V2_Gaussian}
\tilde W^t_{x_t\leftarrow x_{t+\Delta t}}&=& G\left(x_t;x_{t+\Delta t}+\tilde{f}(x_{t+\Delta t})\Delta t ,\sqrt{2\Delta t D(t)}\right)+{\cal O}((\Delta t)^2),
\end{eqnarray}
where
\begin{eqnarray}
\label{eq:AP_Langevin_reversed_transition_prob_V2_Gaussian2}
\tilde{f}(x)&=&-a(t)x+\left(\mu\left(t\right)-x\right)\frac{2D\left(t\right)}{\sigma^{2}(t)}.
\end{eqnarray}
Restoring continuous time, these transition probabilities in turn correspond to the stochastic differential equation for a Gaussian process $\tilde{x}(t)$:
\begin{equation}
\label{eq:Ap_Langevin_reversed_Langevin_eq}
\Dot{\tilde{x}}=-a(t)\tilde{x}+\left[\mu\left(t\right)-\tilde{x}\right]\frac{2D\left(t\right)}{\sigma^{2}(t)}+\sqrt{2D(t)}\xi(t).
\end{equation}
which generalizes previous known results in the literature, e.g. \cite{Chetrite2015}. The drift term for the process $\tilde x$ consists of two contributions. The first, $-a(t)$, is the drift term of the original process [Eq.~(\ref{eq:Ap_Langevin_linear_Langevin_eq})], but with reversed sign.  The second contribution, $\left[\mu\left(t\right)-\tilde{x}\right]\frac{2D\left(t\right)}{\sigma^{2}(t)}$ pulls trajectories towards the mean value $\mu(t)$ with a strength that at time $t$ is inversely proportional to the variance $\sigma^2(t)$ of the original process. The pull becomes infinite as $t\to 0$ given that $\sigma(t=0)= 0$, due to the fixed initial condition $x_0$ of the process $x(t)$. As a consequence all trajectories $\tilde x$ take the value $\tilde x(t=0)=\mu(0)=x_0$.
\medskip

\emph{Example: Brownian motion.} As an example we consider simple Brownian motion,
\begin{equation}
\label{eq:Ap_Langevin_BM}
\Dot{x}=\sqrt{2D}\xi(t),
\end{equation}
with initial condition $x(0)=0$. We then have $\mu(t)=0$ for all $t$, and  $\sigma(t)=\sqrt{2Dt}$. The stochastic differential equation for the associated reverse process $\tilde x$ is 
\begin{equation}
    \label{eq:Ap_Langevin_reversed_BM}
    \Dot{\tilde{x}}=-\frac{\tilde{x}}{t}+\sqrt{2D}\xi(t),
\end{equation}
This equation is to be integrated backwards, that is, from $t=T$ to $t=0$. For any starting point $\tilde{x}(t=T)=x_{_T}$ for this integration, the resulting trajectory $\tilde x$ will necessarily end in $\tilde{x}(t=0)=0$. 

\section{Susceptible-Infected-Susceptible model and WKB method}\label{Ap-SIS}

\subsection{Quasi-stationary distribution and reverse process}
To determine the quasi-stationary distribution of the SIS model we follow the standard approach described for example in \cite{ashcroft2016wkb,grafke2015instanton,grafke2019numerical,Heymann2008,Dembo1998,assaf2010extinction,assaf2017wkb}. We write $x=n/N$ for the fraction of infected individuals in the following. Using the large-deviation ansatz $P(n,t)\propto e^{-N S_0(x,t)}$ in the master equation describing the model and expanding to first order in $\frac{1}{N}$, one obtains a partial differential equation for $S_0(x,t)$ \cite{ashcroft2016wkb},
\begin{eqnarray}\label{eq:AP_SIS_-PDE_for_S_0}
 \frac{\partial S_0(x,t)}{\partial t}=-\sum_{\ell=+1,-1}\frac{\omega_{x\rightarrow x+\frac{\ell}{N}}}{N}\left(e^{\ell\frac{\partial S_0(x,t)}{\partial x}}-1\right).
\end{eqnarray}
Seeking stationary solutions $S_0(x)=\lim_{t\rightarrow \infty}S_0(x,t)$, one finds the following ordinary differential equation for $S_0(x)$ :
\begin{eqnarray}
\frac{d S_0(x)}{d x}=\log \left(\frac{\omega_{x\rightarrow x-\frac{1}{N}}}{\omega_{x\rightarrow x+\frac{1}{N}}}\right).
\end{eqnarray}
Using the rates of the SIS model [Eqs.~\eqref{eq:SIS_rates} in the main paper] this can be solved directly, leading to
\begin{eqnarray}
S_0(x)=x\left(1-\log \, \frac{\beta}{\gamma}\right)+(1-x)\log(1-x)-1+\frac{\gamma}{\beta}+\log\,\frac{\beta}{\gamma},
\end{eqnarray}
where the integration constant was fixed by imposing that $S_0(x)$ is zero at its minimum. The quasi-stationary distribution is then
\begin{eqnarray}\label{eq_PQSSIS}
P^{QS}(x)={\cal N}e^{-NS_0(x)},
\end{eqnarray}
with a suitable normalisation constant ${\cal N}$.

To compute the rates of the reversed process, we use  Eqs.~(\ref{eq:Backward_transition_rates_continuous_time}) after replacing the time-dependent probabilities by the quasi-static distribution in Eq.\eqref{eq_PQSSIS}. We find,
\begin{eqnarray}
\tilde\omega_{n+1 \leftarrow n}&=&\gamma (n+1) \frac{P^\text{QS}(n+1)}{P^\text{QS}(n)},\nonumber\\
\tilde\omega_{n-1\leftarrow n}&=& \beta  \frac{(n-1)(N-n+1)}{N} \frac{P^\text{QS}(n-1)}{P^\text{QS}(n)},
\end{eqnarray} 
These rates were used for the generation of the stochastic trajectories in Fig.~\ref{fig:SIS}.

\subsection{WKB instanton}
Again following along the lines of \cite{ashcroft2016wkb,grafke2015instanton,grafke2019numerical,Heymann2008,Dembo1998,assaf2010extinction,assaf2017wkb}, we introduce $p=\frac{\partial H(x,p)}{\partial x}$ and $H(x,p)=\sum_{\ell=+1,-1}\dfrac{\omega_{x\rightarrow x+\frac{\ell}{N}}}{N}(e^{\ell p}-1)$. 
Eq.~(\ref{eq:AP_SIS_-PDE_for_S_0}) is then recognised as a Hamilton-Jacobi equation
\begin{equation}
    \frac{\partial S_0(x,t)}{\partial t}=-H\left(x,\frac{\partial S_0(x,t)}{\partial x}\right).
    \end{equation}

\begin{figure}
    \centering
    \includegraphics[width=0.5\textwidth]{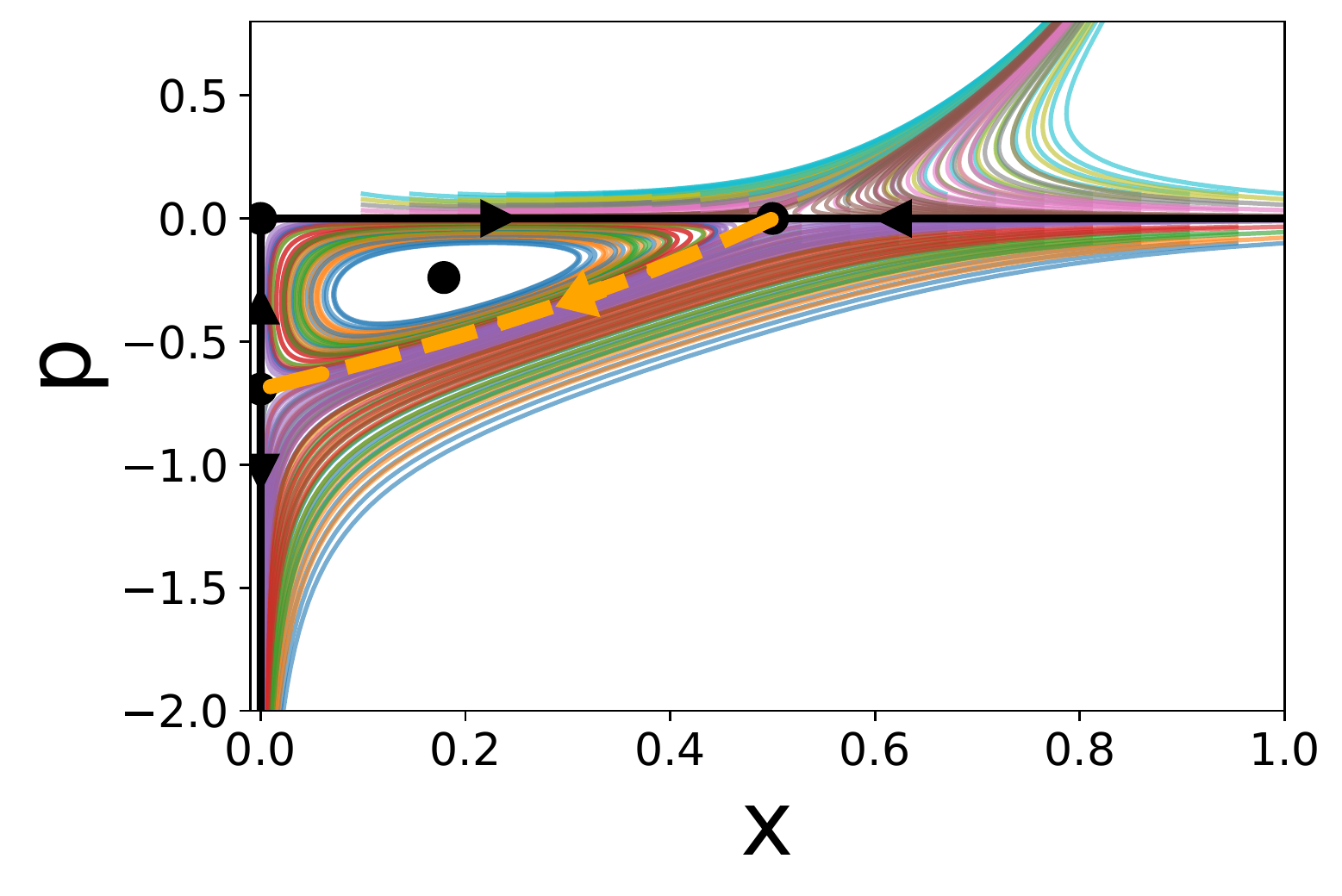}
    \caption{Trajectories of Eqs. (\ref{eq:AP_SIS_Hamilton_equations}) with initial conditions $x(t=0)\in[0.1,1]$ and $p(t=0)\in[-0.1,0.1]$. Dots are placed at the fixed points of Eqs. (\ref{eq:AP_SIS_Hamilton_equations}). The heteroclinic connecting ($x=x_\text{st}, p=0$) with ($x=0,p=\log(\gamma / \beta)$ (dashed line) is the optimal path to extinction plotted in Fig. \ref{fig:SIS}. This heteroclinic is the only solution connecting the line $x=x_{st}$ with the line $x=0$ with all $p(t)$ finite.}
    \label{fig:AP_instantons_SIS}
\end{figure}

Using the rates of the SIS model in Eq.~\eqref{eq:SIS_rates} one obtains 
\begin{eqnarray}
H(x,p)=\beta x(1-x)\left(e^{p}-1\right)+\gamma x \left(e^{-p}-1\right),
\end{eqnarray}
and the associated Hamilton equations 
\begin{align}
\label{eq:AP_SIS_Hamilton_equations}
\dot{x}&=\frac{\partial H(x,p)}{\partial p}=\beta x(1-x)e^{p}-\gamma x e^{-p}, \nonumber\\ 
\dot{p}&=-\frac{\partial H(x,p)}{\partial x}=\beta (2x-1)\left(e^{p}-1\right)-\gamma\left(e^{-p}-1\right).
\end{align}
These equations have four fixed points $(x=0,p=0),\,(x=0,p=\log(\gamma/\beta)),\,(x=x_\text{st},p=0),\,\left(x=\frac{4\beta-\gamma(1+\sqrt{1+8\beta/\gamma})}{8\beta},p=\log(\frac{\gamma(\sqrt{1+8\beta/\gamma}-1}{2\beta})\right)$, with $x_\text{st}\equiv 1-\gamma/\beta$. Fig.~\ref{fig:AP_instantons_SIS} shows the solutions of Eqs.~(\ref{eq:AP_SIS_Hamilton_equations}) in phase space.

Following~\cite{touchette2009large}, we will refer to solutions of Hamilton's equations with two fixed endpoints as instantons. The WKB instanton is the most likely trajectory for the system to take in the limit $N\to \infty$ given any start and end points $x_0$ and $x_{{}_T}$. For the case of the SIS model, it is found by solving Eq.~(\ref{eq:AP_SIS_Hamilton_equations}). 

The instanton connecting ($x=x_\text{st}, p=0$) with ($x=0,p=\log(\gamma / \beta))$ is the dominant path to extinction, shown as a dashed line in Fig. \ref{fig:SIS}a. This trajectory fulfills Eqs.~\eqref{eq:AP_SIS_Hamilton_equations} and $H(x(t),p(t))=0$ (since $H(x=0,p)=0$ and $H$ is conserved along trajectories obeying Hamilton's equations). The constraint $H=0$ leads to
\begin{eqnarray}
\dot{x}=-x\left[(1-x)\beta-\gamma\right],
\end{eqnarray}
with solution
\begin{eqnarray}\label{xtanalytical}
x(t)=\frac{x_\text{st}}{1+e^{\beta x_\text{st}(t-t_0)}}.
\end{eqnarray}
The quantity $t_0$ is an integration constant and can be chosen arbitrarily (the WKB instanton in Fig.~\ref{fig:SIS} is for $t_0=3.178$ so that $x(0) = 0.48$). Here, it is useful to distinguish between two phases of the dynamics of the SIS model. After some initial transient the systems reaches quasi-stationarity in an endemic state. The fraction of infected individuals fluctuates about $x_{\rm st}=1-\gamma/\beta$. This is a long-lived state, the residence time is a random variable with a mean which increases exponentially with the population size $N$. Once the system has left this long-lived state by a random fluctuation, it will approach the absorbing state ($n=0$) on a much shorter time scale of order $N^0$. 

We stress that the WKB instanton in Eq.~(\ref{xtanalytical}) describes the relatively quick transition towards absorption after the system has left the quasi-stationary state. The preceding residence time in the quasi-stationary state is separate and random, and the WKB instanton makes no statement about this time (hence the open parameter $t_0$). The mean residence time in the stationary state can be computed separately\cite{ashcroft2016wkb,wentzell1998random}.  

\subsection{Transition towards extinction}
The most likely path to extinction, obtained in the limit $N\to\infty$, is the heteroclinic orbit in (\ref{xtanalytical}) connecting the fixed points $(x=x_\text{st},p=0)$ and $(x=0,p=\log(\gamma/\beta))$ of Eqs. (\ref{eq:AP_SIS_Hamilton_equations}). This path approaches $x_{\rm st}$ for $t\to -\infty$, and $x=0$ for $t\to\infty$. The typical time scale associated with the transition from the meta-stable state and the absorbing state can however be characterised by the time required for the WKB instanton to reach $x=\epsilon$ starting from  $x=x_\text{st}-\epsilon$. This time difference is independent of the choice of $t_0$ and is easily obtained from Eq.\eqref{xtanalytical} as
\begin{equation}
\tau=\frac{1}{\beta x_\text{st}}\log\left(\frac{x_\text{st}-\epsilon}{\epsilon^2}\right).
\end{equation}
In Fig. (\ref{fig:SIS}b) we have used the value $\epsilon=0.02$, leading to $\tau=7.09$.

The time at which a trajectory of the stochastic SIS model leaves the vicinity of $x_{\rm st}$ and crosses  $x=x_{\rm st}-\epsilon$ is a random variable, determined how long the system resides in the meta-stable state abut $x_{\rm st}$ (see above). The system then moves towards extinction at $x=0$. In order to align these transition paths with the WKB instanton in Eq.\eqref{xtanalytical}, we shift each trajectory in time. More precisely we arbitrarily fix $t_0=3.178$ for the WKB instanton (so that $x(0) = 0.48$), and then choose the time shift for each trajectory of the stochastic systems so as to minimise $\int dt\, [x_{ \rm stochastic}(t)-x_{\rm instanton}(t)]^2$. This is the procedure we used to produce Fig. \ref{fig:SIS} (a) in the main paper. 

\section{Fokker-Planck equation for cell differentiation problem}\label{AP-FP}
The Fokker-Planck equation for the genetic switch model in Eq.~(\ref{eq:Waddington_additive_noise}) is
\begin{eqnarray}
\label{eq:AP_Fokker_Planck}
\partial_t P(x_1,x_2;t)=-\vec{\nabla} \cdot( P(x_1,x_2;t)\vec{F}(x_1,x_2))+ D \nabla^2 P(x_1,x_2;t),
\end{eqnarray}
where $\vec{F}(x_1,x_2)=[f(x_1,x_2),f(x_2,x_1)]$ with $f(u,v)=\frac{u^{n}}{S^{n}+u^2}+\frac{S^n}{S^n+v^n}-u$. We study the system within a square domain $(x_1,x_2)\in [0,L]\times [0,L]$ with reflecting boundary conditions. In Fig. (\ref{fig:cell}b) we used $L=3$. We solved Eq.~(\ref{eq:AP_Fokker_Planck}) numerically.

The numerical integration of Eq.~(\ref{eq:Waddington_additive_noise}) requires a discretisation of space and time. We used a forward time centered space method. The time step and spatial discretisation element need to fulfill the conditions~\cite{toral2014stochastic} $\Delta t < \frac{(\Delta x)^2}{D}$ and $\Delta t < \sqrt{\frac{\Delta x}{|\max_{x_1,x_2}\{\vec{F}(x_1,x_2)\}|}}$. The reflecting boundary conditions are imposed by requiring that the normal component of the probability flux $\vec{J}(x_1,x_2;t)=P(x_1,x_2;t)\vec{F}(x_1,x_2)-D\vec{\nabla} P(x_1,x_2;t)$  vanishes at the boundaries~\cite{gardiner1985handbook}. 
We find that the numerical solution of Eq.~(\ref{eq:AP_Fokker_Planck}) converges to a stationary distribution $P^\text{QS}(x_1,x_2)$ relatively quickly. In Fig. (\ref{fig:cell}b) we used $P(x_1,x_2;T=10|x_1^0=1,x_2^0=1,0)$ as a proxy for the quasi-stationary distribution. We carried out a sensitivity analysis testing different final times $T=10,12,15,20$. This did not result in significant differences for the statistics of paths.

\section{Generation of reverse trajectories for Stochastic Differential Equations}\label{Ap-random_number_generator}
In this section, we describe the strategy to numerically sample paths of the associated bridge process with a target dynamics described by an SDE of the form:
\begin{eqnarray}
\label{eq:AP_RNG_general_SDE}
\dot{x}(t)=f(x,t)+g(x,t)\xi(t).
\end{eqnarray}
Here, the noise $\xi(t)$ has mean zero and correlations $\langle \xi(t)\xi(t')\rangle=\delta(t-t')$, and the SDE is to be interpreted in the It\=o sense. For simplicity, we describe the procedure for a univariate process, but generalization is possible to the multivariate case. Again, the Euler-Maruyama scheme to draw trajectories of the SDE is given by
\begin{eqnarray}
x_{t+\Delta t}=x_t+f(x_t,t)\Delta t +g(x_t,t)\sqrt{\Delta t}\Gamma_t,
\end{eqnarray}
where $\Gamma_t$ are uncorrelated Gaussian variables of mean zero and variance one. The transition rates for this discretisation are 
\begin{eqnarray}
W^t_{x_t\to x_{t+\Delta t}}=G(x_{t+\Delta t};x_t+ f(x_t,t)\Delta t,g(x_t,t)\sqrt{\Delta t}),
\end{eqnarray}
where $G(x;\mu,\sigma)$ is the Gaussian distribution defined in Eq.~\eqref{eq:Ap_Langevin_Gaussian}. This distribution can be easily sampled using, for example, the Box-Muller-Wiener or any other suitable algorithm~\cite{toral2014stochastic}.

Using Eq.~\eqref{eq:Backward_transition_prob} in the main paper, the transition rates for the associated process are
\begin{eqnarray}
\label{eq:AP_RNG_general_SDE2}
\tilde{W}^t_{x_t\leftarrow x_{t+\Delta t}}&=& G\left(x_{t+\Delta t}; x_t+f(x_t,t) \Delta t  ,g(x_t,t)\sqrt{\Delta t}\right)
\times \frac{P(x_t,t)}{P(x_{t+\Delta t},t+\Delta t)}.
\end{eqnarray}
In contrast with the case of a linear SDE [Eq.~\eqref{eq:Ap_Langevin_linear_Langevin_eq}], $G\left(x_{t+\Delta t}; x_t+f(x_t,t) \Delta t  ,g(x_t,t)\sqrt{\Delta t}\right)$ is in general not a Gaussian distribution for $x_t$ (as $x_t$ appears as an argument of the functions $f$ and $g$).

To sample $x_{t}$ from this distribution we use a proposal-rejection technique~\cite{toral2014stochastic}. For the proposal step we approximate  
\begin{eqnarray}
G\left(x_{t+\Delta t}; x_t+f(x_t,t) \Delta t  ,g(x_t,t)\sqrt{\Delta t}\right) &\approx&
G\left(x_{_t};x_{t+\Delta t}-f(x_{t+\Delta t},t+\Delta t)\Delta t ,g(x_{t+\Delta t},t+\Delta t)\sqrt{\Delta t}\right)\nonumber \\
&\equiv& G^\text{prop}(x_t).
\end{eqnarray}
We then draw a proposed value $x_t$ from the Gaussian distribution $G^\text{prop}(x_t)$. This proposal is then accepted with a probability proportional to  $\frac{\tilde{W}^t_{x_t\leftarrow x_{t+\Delta t}}}{G^\text{prop}(x_t)}$.

This method generates values of the random variable $x_t$ distributed according to Eq.~\eqref{eq:AP_RNG_general_SDE2}. There is no further bias or error other than those resulting from the Euler-Maruyama discretisation.

In summary, the proposal-rejection algorithm for the associated reverse process is as follows:\\

If the system is in state $x_{t+\Delta t}=y$ at time $t+\Delta t$ then:
\begin{itemize}
\item[1.] Draw a random number $x$ distributed according to the Gaussian distribution $G\left(x;y-f(y,t+\Delta t) \Delta t,g(y,t+\Delta t)\sqrt{\Delta t}\right)$.
\item[2.] With probability 
    \begin{eqnarray}\label{eq:AP_RNG_h}
    H(x)=C\frac{G\left(y;x+f(x,t) \Delta t,g(x,t)\sqrt{\Delta t}\right)}{G\left(x;y-f(y,t+\Delta t) \Delta t,g(y,t+\Delta t)\sqrt{\Delta t}\right)}\frac{P(x,t)}{P(y,t+\Delta t)}, 
    \end{eqnarray}
   accept $x_t=x$. If rejected, go to step 1. Here, $C$ is chosen such that $\max_x{H(x)}\le 1$.
\end{itemize}
In general, $C\approx 1$ and the average acceptance probability is close to 1.

\newpage
\bibliography{apssamp}

\end{document}